\documentclass[preprint,authoryear,12pt]{elsarticle}
\usepackage{algpseudocode}
\usepackage{algorithm}
\usepackage{amsmath}
\usepackage{graphics}
\usepackage{graphicx}
\usepackage{url}

\newtheorem{thm}{Theorem}[section]
\newtheorem{prop}{Proposition}[section]
\newtheorem{remark}{{\it Remark}}[section]

\newcommand{\la}{\lambda}

\journal{Computational Statistics and Data Analysis}
\begin{document}
\begin{frontmatter}

\title{Optimal feature selection for sparse linear discriminant analysis and its applications in gene expression data}
\author[a,b]{Cheng Wang \corref{c}}
\cortext[c]{Correspondence to: Department of Statistics and Finance, University of Science and Technology of China, Hefei 230026, China. Tel.: +86 551 3603 935.}
\ead{wwcc@mail.ustc.edu.cn}
\author[b]{Longbing Cao}
\author[a] {Baiqi Miao}

\address[a]{Department of Statistics and Finance, University of Science and Technology of China, Hefei, Anhui 230026, China}
\address[b]{Advanced Analytics Institute, University of Technology, Sydney, NSW 2007, Australia}

\begin{abstract}

This work studies the theoretical rules of feature selection in linear discriminant analysis (LDA), and a new feature selection method is proposed for sparse linear discriminant analysis. An $l_1$ minimization method is used to select the important features from which the LDA will be constructed. The asymptotic results of this proposed two-stage LDA (TLDA) are studied, demonstrating that TLDA is an optimal classification rule whose convergence rate is the best compared to existing methods. The experiments on simulated and real datasets are consistent with the theoretical results and show that TLDA performs favorably in comparison with current methods. Overall, TLDA uses a lower minimum number of features or genes than other approaches to achieve a better result with a reduced misclassification rate.
\end{abstract}

\end{frontmatter}

\section{Introduction}
Classification in high-dimensional data is a common problem which has created new challenges for traditional statistical methods. For instance, the classification of leukemia data \citep{golub1999} is a classic high-dimensional example in which there are 7129 genes and 72 samples coming from two classes. Due to the small sample size $n$ and large sample dimension $p$, which are often referred to as ``large $p$, small
$n$'' data, estimators of the sample mean and covariance matrix are usually unstable. In a seminal paper by \cite{bickel2004}, linear discriminant analysis (LDA) was proved to be no better than a random guess when $p/n \to \infty$. In the literature, researchers have proposed two classes of independent rules to deal with high-dimensional classification.

A natural method is to ignore the dependence among the variables and this leads to the so-called naive Bayes classifier, see \cite{dudoit2002} or \cite{bickel2004} for more details. This independent rule has also been well studied in many works such as \cite{dudoit2002a}, \cite{tib2002}, and \cite{barry2005}. However, the correlation ignored by the naive Bayes classifier may be very important for classification. This is partially evidenced by \cite{fan2010road}, who comment that the theoretical misclassification rate of the naive Bayes classifier is higher than that of Fisher's rule unless the true population covariance matrix is diagonal.

An alternative approach involves individual analysis. \cite{fan2008high} proposed using the two-sample $t$-statistic to select features. For every feature, a $t$-score is calculated and the features are chosen by their $t$-scores. Similar rules can also be found in \cite{zuber2009gene}, \cite{tibshirani2006correlation}, and \cite{lai2008genome}. In \cite{fan2008high}, the authors proved that the two-sample $t$-statistic could pick up all the differently expressed features. However, those differently expressed features may not be the best features for classification unless the true population covariance matrix is diagonal. For example, \cite{wu2009} pointed out that in gene analysis, most genes are not expressed sufficiently differently that they can be detected by the $t$-statistic.

\cite{fan2010road} and \cite{mai2012} found that the above rules could result in misleading feature selection and inferior classification based on feature selection by the $t$-statistic or the ignorance of correlations among features. As also pointed out in \cite{wu2009}, there is often a group of correlated genes in gene expression analysis in which correlations cannot be ignored, and the covariance information can help to reduce the misclassification rate. Assuming that the population covariance matrix and mean are sparse, a thresholding procedure is used in \cite{shao2011} to estimate parameters and plug these estimators into the LDA. A constrained $l_1$ minimization method is introduced in \cite{cai2011} to estimate the classification direction, and other methods include those of \cite{wu2009}, \cite{tong2011}, \cite{mai2012}, \cite{fan2010road}, \cite{li2001gene}, and \cite{goeman2004global}.

Just as \cite{fan2008high} commented, the difficulty of high- dimensional classification is intrinsically caused by the existence of many noise features that do not contribute to the reduction of the misclassification rate. Thus, if we can select a subset of important features, the high-dimensional classification will become manageable. In gene expression, especially in diagnostic tests, selecting signature genes for accurate classification is essential \citep{yeung2012predicting}. In this article, we study a theoretical rule to capture the discriminant features for classification. Generally, the best $s$ features for classification are those having the same (or almost the same) theoretical misclassification rate as all $p$ features. When the true linear discriminant 
direction is sparse, we can select a subset of features having the same misclassification rate as all $p$ features. For the asymptotic sparsity situation, the misclassification rate based on our selected features is also close to the theoretical misclassification rate. Our results show that the main condition used in \cite{fan2010road}, \cite{cai2011}, \cite{mai2012}, and \cite{shao2011} ensures that such a small subset of important features which can be selected to derive a more stable and accurate classification result does exist.

In this work, a two-stage LDA (TLDA) is proposed to learn high- dimensional data. TLDA uses $l_1$ minimization, which is a linear program for selecting important features; LDA will then be constructed based on these selected features. Asymptotic results of the proposed TLDA are studied where the consistency and convergence results are given. Experiments show that, under the same regularity conditions as in \cite{fan2010road}, \cite{cai2011}, and \cite{mai2012}, TLDA achieves a better convergence rate. Simulation studies and experiments on real datasets support our theoretical results and demonstrate that TLDA outperforms existing methods.

The rest of the paper is organized as follows. In Section 2, we investigate the theoretical rule of choosing features and the asymptotic results. Evaluations in simulated data are included in Section 3. In Section 4, TLDA is applied to three real datasets to demonstrate its performance on real data. Finally, we conclude the article in Section 5. All the proofs are given in Appendix.

\section{Methods}
Let $X$ be a $p$-dimensional normal random vector belonging to class $k$ if $X \sim N_p(\mu_k,\Sigma),~k=1,2,$ where $\mu_1 \neq \mu_2$, and $\Sigma$ is a positive definite symmetric matrix.
If $\mu_1,\mu_2$, and $\Sigma$ are known, the optimal classification rule is Fisher's linear discriminant rule
\begin{eqnarray} \label{fish}
\delta_F(X)=I\{(X-\mu_a)^T \Sigma^{-1}\mu_d >0\},
\end{eqnarray}
where $\mu_a=(\mu_1+\mu_2)/2,~\mu_d=(\mu_1-\mu_2)/2$, and $I$ denotes the indicator function with value 1 corresponding to classifying $X$ to class 1 and 0 to class 2. Fisher's rule is equivalent to the Bayes rule with equal prior probabilities for two classes. The misclassification rate of the optimal rule is
\begin{eqnarray} \label{fishm}
R=1-\Phi(\Delta_p^{1/2}),~\Delta_p=\mu_d^T\Sigma^{-1}\mu_d,
\end{eqnarray}
where $\Phi$ is the standard normal distribution function.

In practice, Fisher's rule is typically not directly applicable because the parameters are usually unknown and need to be estimated from the samples. Let $\{X_{1,j},j=1,\cdots,n_1\}$ and $\{X_{2,j},j=1,\cdots,n_2\}$ be independent and identically distributed random samples from $N_p(\mu_1,\Sigma)$ and $N_p(\mu_2,\Sigma)$, respectively. The maximum likelihood estimators of $\mu_1,\mu_2,\Sigma$ are
\begin{eqnarray*}
&&\bar{X}_{k}=\frac{1}{n_k}\sum_{j=1}^{n_k} X_{1,j},~~k=1,2,\\
&&S_n=\frac{1}{n}\sum_{k=1}^{2}\sum_{j=1}^{n_k}(X_{k,j}-\bar{X}_k)(X_{k,j}-\bar{X}_k)^T,
\end{eqnarray*}
where $n=n_1+n_2$, and setting
\begin{eqnarray*}
\hat{\mu}_a=\frac{\bar{X}_1+\bar{X}_2}{2},~\hat{\mu}_d=\frac{\bar{X}_1-\bar{X}_2}{2},
\end{eqnarray*}
and $\Sigma^{-1}=S_n^{-1}$ (or generalized inverse $S_n^-$ when $S_n^{-1}$ does not exist), Fisher's rule becomes the classic LDA
\begin{eqnarray*}
\delta_{LDA}(X)=I\{(X-\hat{\mu}_a)^T S_n^{-1}\hat{\mu}_d>0 \},
\end{eqnarray*}
and the misclassification rate of LDA based on sample $\{X_{1,j},j=1,\cdots,n_1\}$ and $\{X_{2,j},j=1,\cdots,n_2\}$ is
\begin{eqnarray*}
R_{LDA}=\frac{1}{2}\Phi(\frac{(\hat{\mu}_a-\mu_1)S_n^{-1} \hat{\mu}_d}{(\hat{\mu}_d^TS_n^{-1}\Sigma S_n^{-1} \hat{\mu}_d)^{1/2}})+\frac{1}{2}\Phi(-\frac{(\hat{\mu}_a-\mu_2)S_n^{-1} \hat{\mu}_d}{(\hat{\mu}_d^TS_n^{-1}\Sigma S_n^{-1} \hat{\mu}_d)^{1/2}}),
\end{eqnarray*}
which has been well studied when $p$ is fixed; more details can be obtained from \cite{anderson2003}.

For classification, the best $s$ features are those with the largest $\Delta_s$, where $\Delta_s$ is the counterpart of $\Delta_p$. We begin with basic notation and definitions. For a vector $a=(a_1,\cdots,a_p)^T$, we define $|a|_0=\sum_{j=1}^p I(a_j \neq 0)$, $|a|_1=\sum_{j=1}^p |a_j|$, and $|a|_2=\sqrt{\sum_{j=1}^p a_j^2}$. For any index set $\mathcal{A} \subset \{1,\cdots,p\}$, $\mathcal{A}^c=\{j \in \{1,\cdots,p\}:j \not\in \mathcal{A}\}$ and $C$ is denoted as a constant which varies from place to place. For any two index sets $\mathcal{A}$ and $\mathcal{A}'$ and matrix $B$, we use $B_{\mathcal{A} \mathcal{A}' }$ to denote the matrix with rows and columns of $B$ indexed by $\mathcal{A}$ and $\mathcal{A}'$. For a vector $b$, $b_{\mathcal{A}}$ denotes a new vector with elements of $b$ indexed by $\mathcal{A}$. In particular, $\Delta_{\mathcal{A}}=(\mu_d)^T_{\mathcal{A}}(\Sigma^{-1})_{\mathcal{A} \mathcal{A}} (\mu_d)_{\mathcal{A}}$, which dominates the theoretical misclassification rate if we only use 
features corresponding to index set $\mathcal{A}$.

The following propositions give solutions to the feature selection problem. Here and below we write $\beta_0=2 \Sigma^{-1}\mu_d$.
\begin{prop} \label{thm1}
Let $\mathcal{A}=\{k:(\beta_0)_k \neq 0\}$. We have
\begin{eqnarray} \label{eq1}
\Delta_{\mathcal{A}}=\mu_d^T\Sigma_p^{-1}\mu_d=\Delta_p.
\end{eqnarray}
\end{prop}
Proposition \ref{thm1} means that the best features are indexed by the support of $\beta_0$. If $\beta_0$ is approximately sparse, which means that many entries of $\beta_0$ are very small, we have the following result.
\begin{prop}\label{thm2}
Assuming that there is a constant $c_0$ (not dependent on $p$) such that $\frac{1}{c_0} \leq all~eigenvalues~of~\Sigma_p \leq c_0$ and there exists $\mathcal{A}_1 \subseteq \{1,2,\cdots,p\}$ satisfying $s_p=\sum_{ k \in \mathcal{A}^c_1} |(\beta_0)_k|^2 \to 0$, we have
\begin{eqnarray}
\Delta_p-\Delta_{\mathcal{A}_1}=O(s_p).
\end{eqnarray}
\end{prop}

Propositions \ref{thm1} and \ref{thm2} provide the theoretical foundations for choosing features, and next we will study how to recover the support of $\beta_0$ from the samples. In other fields, such as compressed sensing and high-dimensional linear regression, constrained $l_1$ minimization has been a common method for reconstructing a sparse signal \citep{donoho2006,candes2007}. In a recent work by \cite{cai2011}, the authors applied $l_1$ minimization to estimate $\beta_0$ directly. However, as \cite{candes2007} pointed out, a two-stage $l_1$ minimization procedure tends to outperform the practical results; more details can be found in the discussions in \cite{candes2007}. Motivated by this, we use $l_1$ minimization in our work to select features and construct LDA on
 those selected features.

First, to ensure the identifiability of the important features, we assume that there exists $\mathcal{A} \subseteq \{1,2,\cdots,p\}$ satisfying $p_0=|\mathcal{A}|_0=o(\sqrt{n/\log{p}})$, $(\beta_0)_{\mathcal{A}^c}=0$, and $\min_{k \in \mathcal{A}}{|(\beta_0)_k|} \geq c_p$. Based on the samples, we first consider the $l_1$ minimization method,
\begin{eqnarray} \label{L1}
\hat{\beta} \in \arg\min_{\beta \in R^p}\{|\beta|_1 ~subject~to~|S_n \beta-(\bar{X}_1-\bar{X}_2)|_{\infty} \leq \la_n\},
\end{eqnarray}
where $\la_n$ is a tuning parameter. Second, important features will be selected as
\begin{eqnarray}
\mathcal{A}^{\ast}=\{j: |\hat{\beta}_j| is~among~the~first~largest~p_0~of~all\} .
\end{eqnarray}
Before introducing the asymptotic properties of TLDA, we specify the following regularity conditions
\begin{eqnarray} \label{maincd}
&&c_0^{-1} \leq n_1/n_2 \leq c_0,~c_0^{-1} \leq \la_{min}(\Sigma_p) \leq \la_{max}(\Sigma_p) \leq c_0, \nonumber \\
&&\log{p} \leq n,~\Delta_p \geq c_0^{-1} ~for~some~constant~ c_0>1, ~~~~~~~~~~~~~~
\end{eqnarray}
which are commonly used in high-dimensional settings. Our first result is the consistency of $\mathcal{A}^{\ast}=\mathcal{A}.$
\begin{thm} \label{thm3}
Let $\la_n=C \sqrt{\Delta_p \log{p}/n}$, with $C>0$ being a sufficiently large constant. Suppose that (\ref{maincd}) holds and that $c_p^2/(\Delta_p p_0 \sqrt{\log{p}/n}) \to \infty$. Then
\begin{eqnarray} \label{mthm3}
P(\mathcal{A}^{\ast}=\mathcal{A})=1-O(p^{-1}).
\end{eqnarray}
\end{thm}
From (\ref{mthm3}), we know that the truly important feature set $\mathcal{A}$ will be indexed by $\mathcal{A}^{\ast}$ with a high probability. If the LDA is constructed on those selected features, the following results demonstrate the explicit convergence rate of the misclassification rate based on features $\mathcal{A}^{\ast}$.
\begin{thm} \label{thm4}
Under the assumption of Theorem \ref{thm3}, and applying LDA to features $\mathcal{A}^{\ast}$, denoting the corresponding misclassification rate as $R_{\mathcal{A}^{\ast}}$, then the following hold.\\
(1) $R_{\mathcal{A}^{\ast}}-R \to 0$ in probability.\\
(2) If further assuming $\Delta_p p_0 \sqrt{\log{p_0}/n} \to 0$,
\begin{eqnarray} \label{crate}
\frac{R_{\mathcal{A}^{\ast}}}{R}-1=O(p_0 \Delta_p \sqrt{\log{p_0}/n}),
\end{eqnarray}
with probability greater than $1-O(p^{-1})$.
\end{thm}
\begin{remark}
According to Definition 1 of \cite{shao2011}, with probability greater than $1-O(p^{-1})$, TLDA is asymptotically optimal when $\Delta_p p_0 \sqrt{\log{p_0}/n} \to 0$. Furthermore, the conditions in Theorems \ref{thm3} and \ref{thm4} are similar to those in \cite{fan2010road}, \cite{mai2012}, and \cite{cai2011}, but our method has a better convergence rate. For example, Theorem 3 in \cite{cai2011} shows that $R_n/R-1=O(p_0 \Delta_p \sqrt{\log{p}/{n}})$. Noting that $p_0\ll p$, therefore our results outperform theirs in this case. This means that, compared with estimating $\beta_0$ directly, our two-stage method  improves the results in theory.
\end{remark}

\section{Simulations}

In practice, the final LDA depends on parameters $\la_n$ which can be selected by maximizing the cross-validation (CV) as in \cite{cai2011} and $p_0$, which can also be selected by CV. Our algorithms are outlined below.
\begin{algorithm}
\caption{A Two-stage LDA based on $l_1$ minimization}
\begin{algorithmic}[1]
\State Calculating the sample covariance matrix $S_n$ and mean $\bar{X}_{k},k=1,2$;
\State  $\hat{\beta}^{\la_n}= \arg\min_{\beta \in R^p} \sum_{k=1}^p |\beta_k| ~subject~to~|S_n \beta-(\bar{X}_1-\bar{X}_2)|_{\infty} \leq \la_n;$
\State   Denoting the tuning parameters chosen by five-fold CV as $\hat{\la}_n$ and $\hat{p}_0$. Here we adjust $\hat{\la}_n$ as $\la=\sqrt{4/5} \hat{\la}_n$;
\State  $\mathcal{A}^{\ast}=\{j: |\hat{\beta}^{\la}_j| is~among~the~first~largest~\hat{p}_0~of~all\}$;
\State  $\beta^{\ast}=((S_n)_{\mathcal{A}^{\ast}\mathcal{A}^{\ast}})^{-1}((\bar{X}_1)_{\mathcal{A}^{\ast}}-(\bar{X}_2)_{\mathcal{A}^{\ast}})$;
\State   If $(Y-(\bar{X}_1+\bar{X}_2)/2)_{\mathcal{A}^{\ast}}^T \beta^{\ast}>0$, classifying $Y$ to class 1, else class 2.
\end{algorithmic}
\end{algorithm}

The reason for adjusting $\hat{\la}_n$ as $\la=\sqrt{4/5} \hat{\la}_n$ is due to $\la_n=C \sqrt{\Delta_p \log{p}/n}$, and the fact that the sample size is $4n/5$ but not $n$ in five-fold CV. The simulations reported in Table 4 of \cite{cai2011} also support our adjustment here. Furthermore, the $l_1$ minimization is a linear program which is very attractive for high-dimensional data and can be implemented by many existing programs, such as the function $linprogPD$ included in the R package ``clime", which is available at \url{http://cran.r-project.org/web/packages/clime/index.html}.

We now present the results of simulation studies which were designed
to evaluate the performance of the proposed TLDA. For the purpose of comparison, we also apply several other methods to the data, specifically, linear programming discriminant (LPD) \citep{cai2011}, regularized optimal affine discriminant (ROAD) \citep{fan2010road,wu2009}, and the oracle Fisher's oracle rule (Oracle). The oracle rule is included as a benchmark. The LPD will be solved by the R package clime and the matlab code for ROAD is available at  \url{http://www.mathworks.com/matlabcentral/fileexchange/40047}.

In simulations, we fix the sample size $n_1=n_2=100$ and without loss of generality we set $\mu_2=0$. For the true classification direction $\beta_0$, $(\beta_0)_{[(2k-1)/10]}=(-1)^{k+1}(k+1)/4,~k=1,\dots,5$ and all other elements are zero. Two kinds of population covariance matrix will be considered.

\begin{itemize}
  \item Model~1. $\Sigma=(\sigma_{ij})_{p \times p}$, where $\sigma_{ij}=0.8^{|i-j|}$ for $1 \leq i,j \leq p$.
  \item Model~2. $\Sigma=(\sigma_{ij})_{p \times p}$, where $\sigma_{ii}=1$ for $1 \leq i \leq p$ and $\sigma_{ij}=0.5$ for $i \neq j$.
\end{itemize}

\begin{figure}
\includegraphics[scale=0.70]{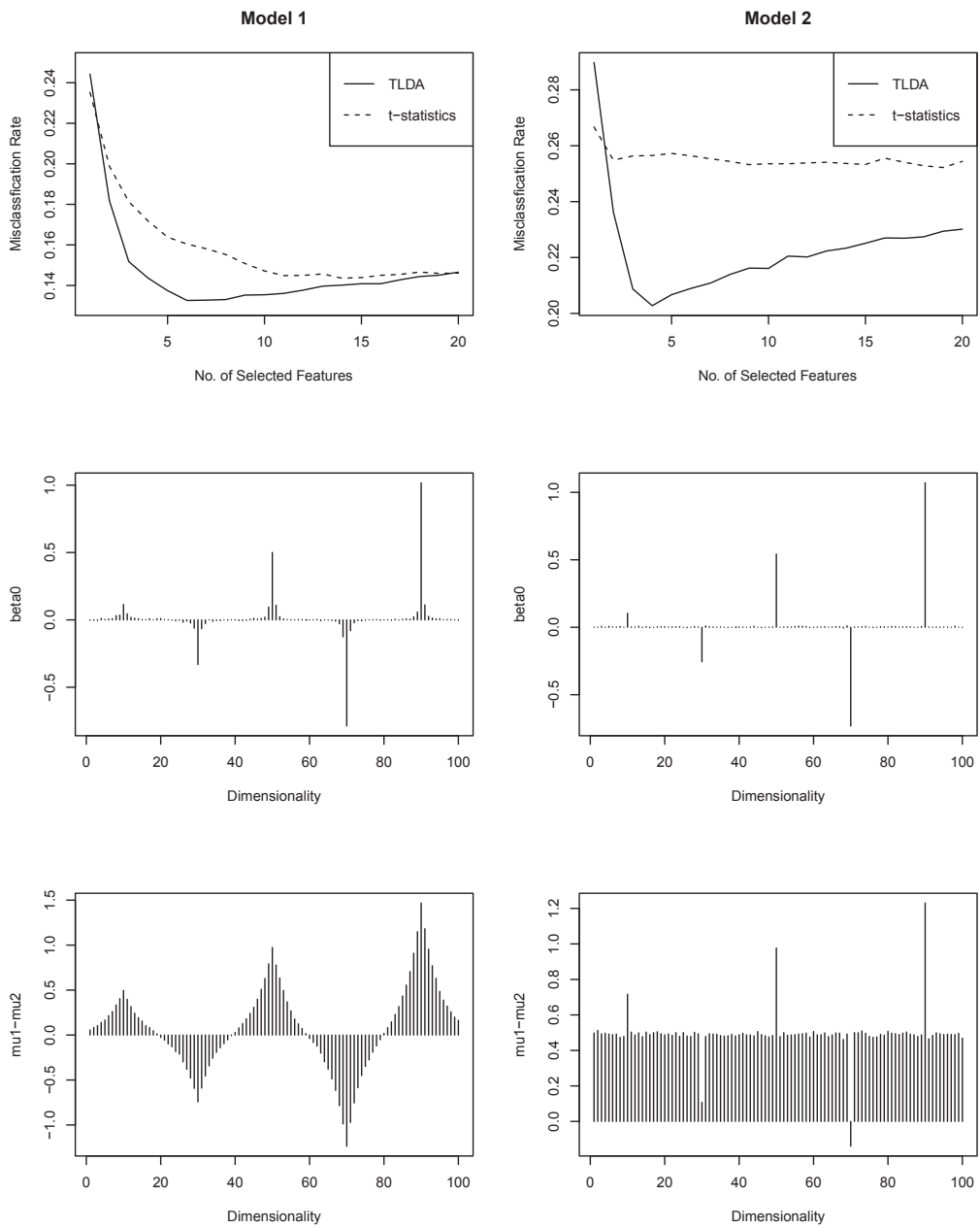}
\caption{Plots for TLDA and the $t$-statistic. Upper: average misclassification rates versus number of selected features; Middle: average $\beta_0$ representing the signal of choosing features by TLDA; Lower: average $\mu_1-\mu_2$ representing the signal of choosing features by the $t$-statistic.}
\label{fig1}
\end{figure}

The first simulation is to evaluate the performance of our proposed TLDA method and the two-sample $t$-statistic \citep{fan2008high}. The average misclassification rates based on 100 simulations are reported in Fig. \ref{fig1}, and here $p=100$.
The figure shows that TLDA always selects more useful features than the two-sample $t$-statistic, which ignores the correlation between features. Specifically, due to correlations, features $30$ and $70$ cannot be detected by the two-sample $t$-statistic for Model 2.

In the second simulation, we study the misclassification rate of our TLDA method. In \cite{cai2011} and \cite{fan2010road}, the authors have conducted many numerical investigations to compare their methods with others, including the oracle features annealed independence rule (OFAIR) \citep{fan2008high} and nearest shrunken centroid (NSC) method \citep{tib2002}, and concluded that their methods perform better. We therefore  compare TLDA only with LPD and ROAD and do not consider other classic methods. Table \ref{table1} shows the misclassification rates based on 100 replications for TLDA, LPD, ROAD, naive Bayes (NB) and Oracle.
\begin{table}[h!]
\caption{Average misclassification rates in percentage for sparse situations. Standard deviations are given in parentheses.}
\label{table1}
\begin{tabular}{cccccc}
\hline
$p$&TLDA&LPD&ROAD&NB&Oracle\\
\hline
\multicolumn{6}{c}{Model~1}\\
100&\textbf{13.41}(2.68)&13.58(2.48)&16.68(5.44)&16.94(2.64)&11.59(2.18)\\
200&\textbf{13.31}(2.45)&13.62(2.55)&16.19(5.05)&17.18(2.54)&11.66(2.38)\\
400&\textbf{13.99}(2.56)&14.06(2.69)&17.45(5.49)&18.86(2.67)&11.88(2.39)\\
800&\textbf{14.16}(2.94)&14.93(2.96)&18.22(5.08)&20.56(2.92)&11.74(2.30)\\
\hline
\multicolumn{6}{c}{Model~2}\\
100&\textbf{20.78}(3.01)&21.04(3.14)&25.01(4.47)&35.13(3.02)&18.41(2.66)\\
200&\textbf{20.91}(3.26)&21.58(3.27)&25.49(3.91)&35.92(2.76)&18.55(2.55)\\
400&\textbf{21.49}(3.50)&22.49(3.55)&26.04(3.88)&35.87(2.86)&18.60(2.76)\\
800&\textbf{21.99}(3.70)&23.31(3.75)&26.62(3.71)&36.04(3.03)&18.70(3.13)\\
\hline
\end{tabular}
\end{table}

From Table \ref{table1}, we can see that the performance of TLDA is similar to that of Oracle and is better than that of the other methods. Clearly, due to its fundamental drawback, the naive Bayes is the worst of all methods although it is better than random guess (whose misclassification rate is 50\%). Overall, compared with LPD and ROAD, TLDA has the smallest misclassification rate, and the standard deviation of TLDA is similar to that of LPD but smaller than that of ROAD. When the dimensionality $p$ increases from 100 to 800, TLDA is quite stable, whereas LPD and ROAD become increasingly worse. In particular, TLDA always has a smaller misclassification rate and standard deviation than ROAD. When $p$ is not large, TLDA and LPD have similar performance, while TLDA becomes better than LPD as $p$ increases; in particular when $p$ is sufficiently large (such as $p=800$), the difference between the misclassification rates of TLDA and LPD becomes bigger. In summary, simulations demonstrate that TLDA is a stable and superior classification method 
compared to existing methods.

\begin{figure}
\centering
\includegraphics[scale=0.6]{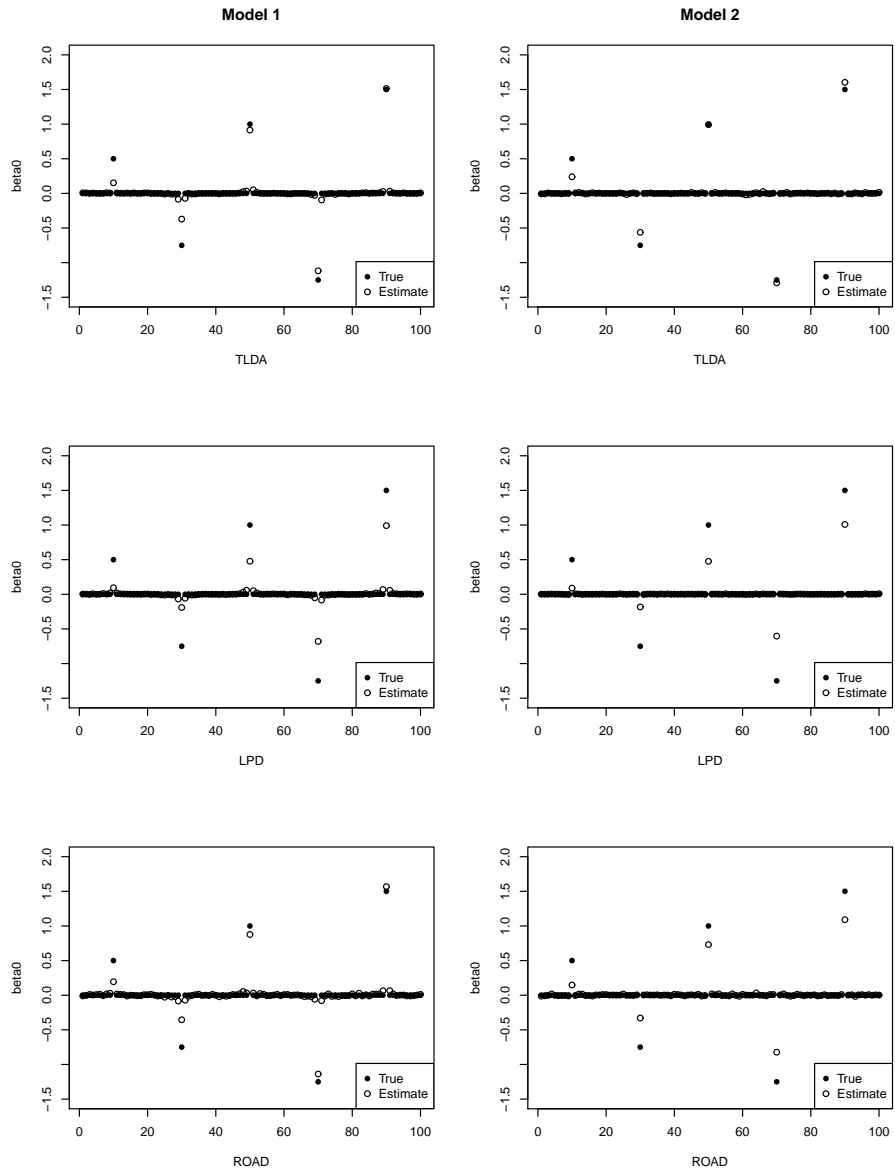}
\caption{Average estimators of TLDA, LPD and ROAD for $p=100$. The true $\beta_0$ and the estimators are very sparse, which is why there is an almost solid line at zero.}
\label{fig2}
\end{figure}

Next, we will study the estimators $\hat{\beta}_{TLDA},\hat{\beta}_{LPD}$, and $\hat{\beta}_{ROAD}$. Fig. \ref{fig2} plots the average estimators of 100 replications. Due to different assumptions, here we adjust $\hat{\beta}_{ROAD}$ to $|\beta_0|^2*\hat{\beta}_{ROAD}$ so that it fits the real situation.
From Fig. \ref{fig2}, we can see that TLDA correctly selects most of those five features but very few noise features. In particular, compared with LPD, which estimates the true $\beta_0$ directly, our two-stage estimators are much closer to $\beta_0$, which is consistent with the  discussions in \cite{candes2007}.

The above simulations are conducted for scenarios where $\beta_0$ is sparse. In practice, it is quite common that there are many weak signals that are correlated with the main signals. It would be interesting to evaluate the performance of TLDA for these approximately sparse situations.  Specifically, we will consider two scenarios with respect to $\mu_1$, as  follows.
\begin{itemize}
  \item Model 3. $\mu_1=(1_5,0_{p-5})$ in Model 1.
  \item Model 4. $\beta_0=0.551*(3,1.7,-2.2,-2.1,2.55,(p-5)^{-1}1_{p-5})$ and $\mu_1=\Sigma*\beta_0$ in Model 2.
\end{itemize}   
Here $n_1=n_2=100$ and $\mu_2=0$. Model 3 is similar to those in \cite{cai2011} and \cite{fan2010road}, and Model 4 comes from \cite{mai2012}. The average misclassification rates based on 100 replications are reported in Table \ref{tablen1}. It is again evident that TLDA performs favorably compared to existing
methods.

\begin{table}[h!]
\caption{Average misclassification rates in percentage for approximately sparse simulations. Standard deviations are given in parentheses.}
\label{tablen1}
\begin{tabular}{cccccc}
\hline
$p$&TLDA&LPD&ROAD&NB&Oracle\\
\hline
\multicolumn{6}{c}{Model 3}\\
100&\textbf{20.70}(3.12)&22.69(3.67)&26.85(5.91)&31.46(4.07)&18.56(2.54)\\
200&\textbf{20.89}(3.11)&24.03(3.83)&27.52(5.37)&33.74(3.68)&18.98(2.65)\\
400&\textbf{20.96}(3.18)&25.03(3.77)&28.03(5.36)&36.61(3.69)&18.65(2.59)\\
800&\textbf{21.75}(4.56)&26.77(4.60)&28.73(5.14)&40.71(3.63)&18.80(2.69)\\
\hline
\multicolumn{6}{c}{Model 4}\\
100&\textbf{11.99}(2.68)&12.30(2.59)&14.57(3.33)&21.87(2.68)&9.98(2.07)\\
200&\textbf{12.64}(2.58)&13.04(2.67)&15.15(3.19)&22.17(2.97)&10.60(2.06)\\
400&\textbf{12.70}(2.64)&13.52(2.40)&15.56(3.09)&22.28(2.79)&10.03(2.17)\\
800&\textbf{12.90}(3.01)&13.85(2.94)&15.35(3.75)&22.33(3.11)&10.08(2.21)\\
\hline
\end{tabular}
\end{table}

\section{Real data}
In this section, we apply the proposed TLDA to real datasets. Since real data usually has an ultra-high data dimension $p$, a sure independence screening (SIS) method \citep{fan2008sure} will be carried out before our proposed feature selection procedure to further improve the accuracy and control the computational cost. For brevity, we will apply the two-sample $t$-test statistic \citep{tib2002, fan2008high} to reduce the dimensionality from ultra-high to a moderate scale. Other screening steps such as that in \cite{fan2010road} can also be used, but we do not pursue them in detail.  

First, TLDA is applied to study leukemia data, which is available at \url{http://www.broadinstitute.org/cgi-bin/cancer/datasets.cgi.} The dataset contains $p=7129$ genes for $n_1=27$ acute lymphoblastic leukemia (ALL) samples and $n_2=11$ acute myeloid leukemia (AML) samples in the training set; the test set consists of 20 ALL samples and 14 AML samples. More details can be found in \cite{golub1999}. By following similar pre-processing steps as \cite{dudoit2002} and \cite{fan2008high}, we standardize each sample to zero mean and
$S_n=\frac{1}{n}\sum_{k=1}^{2}\sum_{j=1}^{n_k}(X_{k,j}-\bar{X}_k)(X_{k,j}-\bar{X}_k)^T$ has unit diagonal elements.

\begin{figure}[h!]
\centering
\includegraphics[scale=0.80]{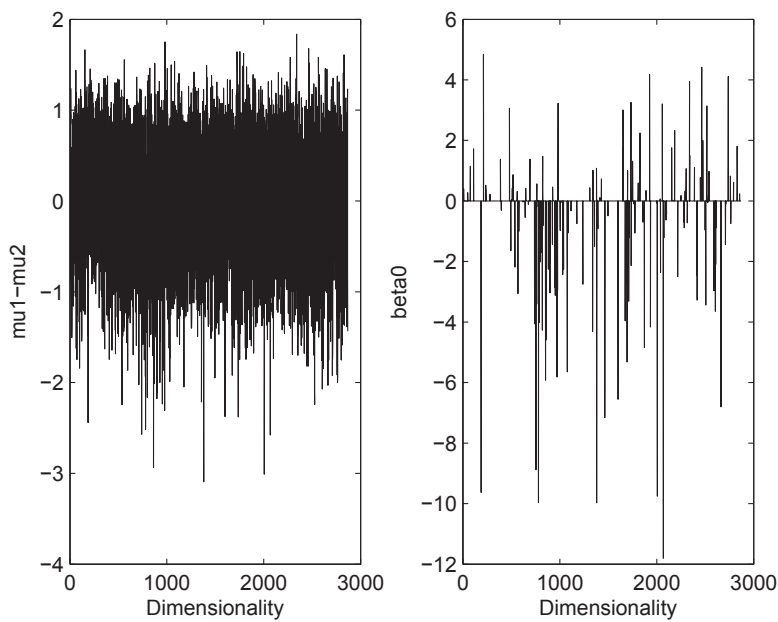}
\caption{True mean difference and estimator $\hat{\beta}_0$ of leukemia data.}
\label{fig3}
\end{figure}

For comparison with LPD in \cite{cai2011}, we use 2867 genes with the largest absolute values of the two-sample $t$-statistic ($|\mu_1-\mu_2|>0.5$). Fig. \ref{fig3} shows the mean difference and estimator $\hat{\beta}_0$ (tuning parameter $\la=1.2$), representing the feature selection signals of the two-sample $t$-statistic and TLDA, respectively.
Clearly, the signal for  TLDA is sparse, while the signal for the two-sample $t$-statistic has no clear clues.
The classification results for TLDA, LPD, ROAD, OFAIR, NSC, and NB are shown in Table \ref{table3}.
\begin{table}
\caption{Classification errors of leukemia data by various methods}
\label{table3}
\begin{tabular}{ccccccc}
\hline
 &TLDA&LPD&ROAD&OFAIR&NSC&NB\\
\hline
Training~Error&0/38&0/38&0/38&1/38&1/38&0/38\\
Test~Error&1/34&1/34&1/34&1/34&3/34&5/34\\
No. of Selected genes&8&151&40&11&24&7129\\
\hline
\end{tabular}
\end{table}

Table \ref{table3} shows that TLDA performs competitively in classification error with LPD and ROAD. However, TLDA only selects 8 genes, in contrast to 40 genes by ROAD and 151 genes by LPD. The 8 selected genes and their TLDA weights are given in Table \ref{table4}. For comparison, we also present their $t$-statistic rank in the 7129 genes. 

\begin{table}
\centering
\caption{The eight genes of leukemia data selected by TLDA.}
\label{table4}
\begin{tabular}{cccccc}
\hline
Gene position & & TLDA weights& &Rank of $t$-statistic\\
\hline
461& &-3.203& &7\\
1779& &-4.455& &87\\
1834& & -5.039& &6\\
3320& & -0.960& &1\\
3525& &-3.876& &138\\
4847& &-6.389& &2\\
5039& &-1.187& & 4\\
6539& &-7.9933& &21\\
\hline
\end{tabular}
\end{table}

We further compare the methods on two more real datasets: the colon \citep{srivastava2007comparison} and breast cancer \citep{hess2006pharmacogenomic} datasets. A leave-one-out cross validation (LOOCV) is performed on the two datasets. For $i = 1, \cdots, n$ , the $p \times 1$ vector $x_i$ is treated as the testing set, while the remaining $n-1$ observations form the training set.  A subset of 1000 genes is selected based on the two-sample t-statistic. The classification results for the TLDA, LPD, ROAD, and NB methods are shown in Table \ref{table5}. We can see that, on each dataset, the proposed TLDA has a competitive performance in terms of classification errors while using the fewest genes. Overall, 
TLDA is also applicable in real datasets and performs favorably in comparison to existing methods.  
\begin{table}[h!]
\centering
\caption{Classification error and number of genes selected by various methods for the colon and breast cancer datasets}
\label{table5}
\begin{tabular}{cccccc}
\hline
& & TLDA & LPD& ROAD&NB\\
\hline
Colon&  Error(\%)&9.68&9.68&11.29&14.52\\
 & No. of genes & 7.42(1.03)& 168.95(71.39)& 38.10(27.60)& 1000(0)\\
Breast & Error(\%)&21.80&25.56&31.58&34.59\\
& No. of genes & 14.61(2.40)& 332.45(103.56)& 44.14(47.26)& 1000(0)\\ 
\hline   
\end{tabular}
\end{table}
\section{Discussions}
In this paper, we have proposed a solution for feature selection in high-dimensional data.
We have derived the optimal feature selection rule for LDA and proposed the selection of features based on the sparsity of $\Sigma^{-1}\mu_d$. An $l_1$ minimization method is used on the samples to select the important features and LDA is then applied to those selected features. Our proposed TLDA performs favorably compared to existing methods in theory and application. Our analysis shows that the independent rules such as the two-sample $t$-statistic and naive Bayes may not be efficient and may even lead to bad classifiers.

Suppose that there are $K>2$ classes (in this article we assume that $K=2$), our TLDA is also applicable. For this, $X$ will be classified to class $k$ if and only if
\begin{eqnarray}
(X-(\bar{X}_k+\bar{X}_l)/2)_{\mathcal{A}_{kl}^{\ast}}^T \beta_{kl}^{\ast}>0~for~all~k \neq l.
\end{eqnarray}
Moreover, the procedure can be extended to unequal prior probabilities $\pi_1$ and $\pi_2$ in which we classify $X$ to class 1 when
\begin{eqnarray}
(X-(\bar{X}_1+\bar{X}_2)/2)_{\mathcal{A}^{\ast}}^T \beta^{\ast}>\log{(\pi_2/\pi_1)},
\end{eqnarray}
where the parameters can also be estimated as $\hat{\pi}_1=n_1/n$ and $\hat{\pi}_2=n_2/n$. For non-Gaussian distributions, we can also derive similar results under the moment conditions, as in \cite{cai2011}.

Finally, we note that the number of selected features is $p_0=o(\sqrt{n /\log{p}})$ which is very small compared to $p$. Setting $n=O((\log{p})^\beta)$ for  $\beta>1$, this means that only $o((\log{p})^{(\beta-1)/2})$ features can be selected from $p$ variables to apply LDA. This is due to the fact that LDA is stable only when $p_0 \sqrt{p_0/n} \to 0$, and a detailed result can be found in \cite{shao2011}. Our future research will focus on improving $p_0$.

\section*{Acknowledgments}
We are grateful for the valuable comments from the reviewers and editors. Cheng Wang's research was supported by NSF of China Grants (No. 11101397, 71001095 and 11271347). Longbing Cao's research was supported by  Australian Research Council Discovery Grants (DP1096218 and DP1301691) and Australian Research Council Linkage Grant (LP100200774).
\section*{Appendix A: Proofs}

\subsection*{A.1. Proof of Theorem \ref{thm3}}
From the proofs of Theorem 2 in \cite{cai2011}, we know that
\begin{eqnarray} \label{ap1}
(\hat{\beta}-\beta_0)^T \Sigma (\hat{\beta}-\beta_0) \leq C |\beta_0|_1^2 \sqrt{\log{p}/n}+6 \la_n |\beta_0|_1,
\end{eqnarray}
with probability greater than $1-O(p^{-1})$. Using the Cauchy-Schwartz inequality,
\begin{eqnarray*}
&&|\beta_0|_1^2 \leq |\beta_0|_0 |\beta_0|_2^2 \leq c_0 p_0 (\beta_0^T \Sigma \beta_0)= 4 c_0 p_0 \Delta_p ,\\
&& (\hat{\beta}-\beta_0)^T \Sigma (\hat{\beta}-\beta_0) \geq c_0^{-1} (\hat{\beta}-\beta_0)^T (\hat{\beta}-\beta_0).
\end{eqnarray*}
Together with (\ref{ap1}), we have
\begin{eqnarray}
(\hat{\beta}-\beta_0)^T (\hat{\beta}-\beta_0) \leq C p_0\Delta_p \sqrt{\log{p}/n},
\end{eqnarray}
with probability greater than $1-O(p^{-1})$. For $j \in \mathcal{A}$,
\begin{eqnarray*}
|\hat{\beta}_j-(\beta_0)_j|^2 \leq C p_0\Delta_p \sqrt{\log{p}/n}.
\end{eqnarray*}
Then
\begin{eqnarray*}
|\hat{\beta}_j| &\geq& |(\beta_0)_j|-\sqrt{C p_0\Delta_p \sqrt{\log{p}/n}}\\
 &\geq& c_p(1-\sqrt{C p_0\Delta_p \sqrt{\log{p}/n}}/c_p )\\
 &>& c_p/2.
\end{eqnarray*}
Similarly, for $j \in \mathcal{A}^c$,
\begin{eqnarray*}
|\hat{\beta}_j| \leq \sqrt{C p_0\Delta_p \sqrt{\log{p}/n}} < c_p/2.
\end{eqnarray*}
Hence, we have proved that $P(\mathcal{A}^{\ast}=\mathcal{A})=1-O(p^{-1})$.
\subsection*{A.2. Proof of Theorem \ref{thm4}}
Applying the features selector $\mathcal{A}^{\ast}$ to the sample $\{X_{1,j},j=1,\cdots,n_1\}$ and $\{X_{2,j},j=1,\cdots,n_2\}$, we still denote the corresponding data as $X, \{X_{k,j},k=1,2\}$ for brevity. It is noted that here the dimension is $p_0$ not $p$. Setting
\begin{eqnarray*}
&&\bar{X}_{k}=\frac{1}{n_k}\sum_{j=1}^{n_k} X_{1,j},~~k=1,2,\\
&&S_n=\frac{1}{n}\sum_{k=1}^{2}\sum_{j=1}^{n_k}(X_{k,j}-\bar{X}_k)(X_{k,j}-\bar{X}_k)^T,
\end{eqnarray*}
and
\begin{eqnarray*}
\hat{\mu}_a=\frac{\bar{X}_1+\bar{X}_2}{2},~\hat{\mu}_d=\frac{\bar{X}_1-\bar{X}_2}{2}.
\end{eqnarray*}
The LDA procedure is
\begin{eqnarray*}
\delta_{LDA}(X)=I\{(X-\hat{\mu}_a)^T S_n^{-1}\hat{\mu}_d \},
\end{eqnarray*}
and the misclassification rate is
\begin{eqnarray*}
R_{\mathcal{A}^{\ast}}=\frac{1}{2}\Phi(\frac{(\hat{\mu}_a-\mu_1)S_n^{-1} \hat{\mu}_d}{(\hat{\mu}_d^TS_n^{-1}\Sigma S_n^{-1} \hat{\mu}_d)^{1/2}})+\frac{1}{2}\Phi(-\frac{(\hat{\mu}_a-\mu_2)S_n^{-1} \hat{\mu}_d}{(\hat{\mu}_d^TS_n^{-1}\Sigma S_n^{-1} \hat{\mu}_d)^{1/2}}).
\end{eqnarray*}
By the proofs of Theorem 1 in \cite{shao2011}, we know that
\begin{eqnarray*}
\frac{(\hat{\mu}_a-\mu_1)S_n^{-1} \hat{\mu}_d}{(\hat{\mu}_d^TS_n^{-1}\Sigma S_n^{-1} \hat{\mu}_d)^{1/2}}=-\Delta_p^{1/2}(1+O(p_0 \sqrt{\log{p_0}/n})),
\end{eqnarray*}
and a similar result also holds for $\Phi(\frac{(\hat{\mu}_a-\mu_1)S_n^{-1} \hat{\mu}_d}{(\hat{\mu}_d^TS_n^{-1}\Sigma S_n^{-1} \hat{\mu}_d)^{1/2}})$. Then
\begin{eqnarray}
 R_{\mathcal{A}^{\ast}}=\Phi(-\Delta_p^{1/2}(1+O(p_0 \sqrt{\log{p_0}/n}))).
\end{eqnarray}
Noting that $p_0 \sqrt{\log{p_0}/n} \to 0$, therefore, in probability,
\begin{eqnarray}
R_{\mathcal{A}^{\ast}}-R \to 0.
\end{eqnarray}
From equation (12) of \cite{cai2011}, we know that
\begin{eqnarray*}
|\frac{\Phi(\frac{(\hat{\mu}_a-\mu_1)S_n^{-1} \hat{\mu}_d}{(\hat{\mu}_d^TS_n^{-1}\Sigma S_n^{-1} \hat{\mu}_d)^{1/2}})}{\Phi(-\Delta_p^{1/2})}-1|
\leq O(\Delta_p p_0 \sqrt{\log{p_0}/n})e^{O(\Delta_p p_0 \sqrt{\log{p_0}/n})}.
\end{eqnarray*}
Then
\begin{eqnarray*}
|\frac{R_{\mathcal{A}^{\ast}}}{R}-1|\leq O(\Delta_p p_0 \sqrt{\log{p_0}/n})e^{O(\Delta_p p_0 \sqrt{\log{p_0}/n})}.
\end{eqnarray*}
When $\Delta_p p_0 \sqrt{\log{p_0}/n} \to 0$, we get
\begin{eqnarray}
|\frac{R_{\mathcal{A}^{\ast}}}{R}-1|=O(\Delta_p p_0 \sqrt{\log{p_0}/n}).
\end{eqnarray}
The proof is completed.

\bibliographystyle{elsarticle-harv}

\bibliography{cit}
\end{document}